\documentclass
[aps,pra,amsfonts,amssymb,twocolumn,amsmath,preprintnumbers,nofootinbib,floatfix,
showpacs]{revtex4-1}%
\usepackage[dvips]{graphics}
\usepackage{graphicx}
\usepackage{bm}
\usepackage{amsmath}
\usepackage{amsfonts}
\usepackage{amssymb}
\usepackage{xcolor}
\usepackage{subfigure}
\usepackage{hyperref,hypcap}
\usepackage{braket}
\usepackage{commath}%
\providecommand{\U}[1]{\protect\rule{.1in}{.1in}}

\begin{document}

\title{Boltzmann transport from density matrix theory: interband and intraband coherences}
\author{Cong Xiao, Jihang Zhu, and Bangguo Xiong}
\affiliation{Department of Physics, The University of Texas at Austin, Austin, Texas 78712, USA}

\begin{abstract}
To account for the anomalous/spin Hall conductivities and spin-orbit
torque in the zeroth order of electron scattering time in strongly
spin-orbit coupled systems, the Boltzmann transport theory in the case
of weak disorder-potentials has been augmented by adding
some interband coherence effects by hand. In this work these interband
coherence terms are derived systematically from analyzing
the equation of motion of the single-particle density matrix in the Bloch
representation. Interband elements of the out-of-equilibrium density matrix
are related to only one part of interband-coherence responses.
Disorder-induced off-diagonal elements of the equilibrium density matrix are
shown to be vital in producing the coordinate-shift anomalous driving term in
the modified Boltzmann equation. Moreover, intraband coherence is inherent in
the Boltzmann equation, whose contribution to anomalous/spin Hall
conductivities is parametrically the same as the interband coherence.
\end{abstract}
\maketitle

\section{Introduction}

The Boltzmann transport theory has been generally accepted as a qualitatively
good and intuitive starting point in discussing nonequilibrium phenomena in
weakly disordered crystals \cite{Ziman1960}, including recent focuses on the
anomalous Hall effect \cite{Sinitsyn2008,Nagaosa2010}, spin Hall effect
\cite{Sinova2015,Zhang2005,Xiao2017SHE,Xiao2018JPCM}, valley Hall effect
\cite{Xiao2007} and spin-orbit torque \cite{Lee2015,Xiao2017SOT-SBE}. In
applications to these phenomena, the coherence between Bloch states in
different bands, i.e., interband coherence, caused by both the electric field
and disorder \cite{Sinitsyn2008,Nagaosa2010,Kovalev2010} has been incorporated
via semiclassical constructions (Berry-curvature anomalous velocity
\cite{Xiao2010} and scattering-induced coordinate-shift
\cite{Sinitsyn2006,Sinitsyn2007}) or semi-phenomenological arguments
(scattering-induced interband-coherence dressing of carrier states
\cite{Xiao2017SOT-SBE,Xiao2017SHE}). The resultant contribution to the
linear-response coefficients in aforementioned nonequilibrium phenomena is of
the zeroth order of the electron scattering time, and experiments usually
suggest the dominance of\ this sort of contribution in moderately dirty
transition metal samples \cite{Miyasato2007,He2012,Sagasta2016,Buhrman2016}.

Despite the practical success, microscopic understanding of this augmented
Boltzmann formalism has not been complete. Although Kohn and Luttinger have
laid the foundation for the Boltzmann formalism of the anomalous Hall effect
on a density matrix perturbation theory in the case of weak impurity
potentials sixty years ago \cite{KL1957,Luttinger1958}, their classical paper
is too complicated to be absorbed by most researchers in the modern community
of spin and valley Hall effects and spin-orbit torque. Kohn and Luttinger, in
the very early years of modern transport theories, aimed to lay a foundation
not only for the Boltzmann theory of electrical conductivity but also for the
whole metallic conduction theory. From the modern point of view, this aim is
beyond the scope of the Kohn-Luttinger density matrix approach, which only
works in limited parameter regime and suffers from the lack of a systematic
renormalization scheme \cite{Moore1967,Watabe1966}. Therefore, this approach
is much less employed than other quantum transport approaches based on Green's
functions (e.g., Kubo-Streda and Keldysh \cite{Nagaosa2010,Sinova2015}), into
which systematic renormalization procedures can be incorporated.

On the other hand, if one only aims to derive, in the case of weak
disorder-potentials, the Boltzmann formalism which was born in the Bloch
representation of the disorder-free equilibrium single-particle Hamiltonian,
then the density matrix approach is still the most intuitive starting point.
In fact, when constructing the modified Boltzmann theory of anomalous Hall
effect, Sinitsyn $et$ $al.$ \cite{Sinitsyn2006} noticed the correspondence
between the semiclassical coordinate-shift effects and the sum of some
gauge-dependent equations of Luttinger \cite{Luttinger1958}. However, the
Boltzmann theory for the anomalous Hall effect cannot be directly used to
account for spin Hall effect (when the spin is not conserved due to strong
band-structure spin-orbit coupling) and spin-orbit torque. Recently Xiao $et$
$al.$ have argued that the scattering-induced interband-coherence dressing of
carrier states \cite{Xiao2017SOT-SBE,Xiao2017SHE} contributes to the spin Hall
conductivity and spin-orbit torque, playing the role of the side-jump velocity
in the anomalous Hall effect. But this semi-phenomenological construction has
not been confirmed by the density matrix theory.

In this work we show that, all the aforementioned interband-coherence terms
added-by-hand in the Boltzmann formalism can be derived systematically from a
density matrix perturbation analysis with respect to the weak
disorder-potential \cite{KL1957,Luttinger1958}, in the case of Bloch electrons
in non-degenerate multiple-bands scattered by weak Gaussian static disorder.

In particular, the disorder-dependent part of the interband elements of the
\textit{out-of-equilibrium} density-matrix leads to the disorder-induced
interband-coherence dressing of Bloch states in the Boltzmann theory
\cite{Xiao2017SOT-SBE,Xiao2017SHE}. The latter is the only part of
interband coherence responses that arises from interband elements
of the out-of-equilibrium density matrix. We emphasize this because it is
sometimes misunderstood that the conventional Boltzmann formalism \cite{Ziman1960}
only misses the interband elements of the out-of-equilibrium density matrix.
In fact, the combination of the disorder-induced off-diagonal (in the Bloch representation $|l\rangle
=|\eta\mathbf{k}\rangle$ with $\eta$ the band index and $\mathbf{k}$ the
momentum) elements of the \textit{equilibrium} density matrix and the diagonal
perturbations by the electric field leads to the coordinate-shift induced
anomalous driving term in the modified Boltzmann equation \cite{Xiao2017AHE}.

The interband coherence process also occurs in the conventional Boltzmann
equation as long as the scattering amplitude is calculated up to the second
Born order \cite{Sinitsyn2007,Sinitsyn2008}. This so-called
intrinsic-skew-scattering contribution \cite{Sinitsyn2007} arises from
asymmetric differential scattering cross-section on rare impurity pairs
separated by distances of the order of the Fermi wavelength
\cite{Ado2015,Ado2016,Ado2017,Konig2016}. Therefore, contributions to the
differential cross-section from both crossed and noncrossed impurity-lines of
this two-impurity complexes are parametrically the same
\cite{Ado2015,Konig2016}. Previous Boltzmann theories
\cite{Sinitsyn2007,Sinitsyn2008,Xiao2017AHE} only addressed the noncrossed
contribution which involves an interband virtual scattering process. In this
work we show explicitly that the crossed contribution involves not only
interband but also intraband virtual scattering processes. The latter means
coherence between Bloch states in the same band but with different energies,
and an intermediate state of the virtual process lies away from the Fermi
surface (off-shell).

The above main ideas are analyzed in Sec. II, with the main results given in
Eqs. (\ref{density matrix}) -- (\ref{rho-1}), Eq. (\ref{diagonal f}) and Eqs.
(\ref{Culcer}) -- (\ref{c}). The paper is concluded by some discussions in
Sec. III. Some calculation details are presented in the Appendix.

\section{Derivation and Analysis}

\subsection{Preliminaries: Density matrix approach}

In this subsection we just outline the basic framework of the density matrix
equation-of-motion approach proposed by Kohn and Luttinger
\cite{KL1957,Luttinger1958} in the case of weak-potential static disorder.

We introduce the notation $\tilde{A}$ to stand for the representation of
operator $\hat{A}$ in the second-quantized formalism. For a single-carrier
operator, i.e., $\hat{A}=\sum_{i}\hat{A}_{i}$ where $\hat{A}_{i}$ depends only
on the dynamical variables of the $i$-th carrier, one has $\tilde{A}%
=\sum_{nn^{\prime}}A_{nn^{\prime}}a_{n}^{\dag}a_{n^{\prime}}$ where
$A_{nn^{\prime}}$ are the matrix elements in the $n$ representation of
single-carrier space, $a_{n}^{\dag}$ ($a_{n}$) is the creation (annihilation)
operator on the single-carrier eigenstate $\left\vert n\right\rangle $. The
expectation value of $\hat{A}$ is given by $\left\langle A\right\rangle
=Tr\left(  \tilde{\rho}_{T}\tilde{A}\right)  $, where $Tr$ denotes the trace
operation in the occupation-number space, and the many-particle density matrix
$\tilde{\rho}_{T}$ in the occupation-number representation is governed by the
quantum Liouville equation $i\hbar\frac{\partial}{\partial t}\tilde{\rho}%
_{T}=\left[  \tilde{H}_{T},\tilde{\rho}_{T}\right]  $. The expectation value
of a single-carrier operator $\hat{A}$ can then be expressed in terms of
$\hat{A}$ and a single-carrier operator $\hat{\rho}_{T}$:
\begin{align}
\left\langle A\right\rangle  &  =\sum_{nn^{\prime}}A_{nn^{\prime}}\left(
\hat{\rho}_{T}\right)  _{n^{\prime}n}=tr\left[  \hat{A}\hat{\rho}_{T}\right]
,\text{ }\nonumber\\
\left(  \hat{\rho}_{T}\right)  _{n^{\prime}n}  &  \equiv Tr\left(  \tilde
{\rho}_{T}a_{n}^{\dag}a_{n^{\prime}}\right)  .
\end{align}
Here $tr$ denotes the trace in single-carrier Hilbert space.

As Kohn and Luttinger have noticed \cite{KL1957}, when the total Hamiltonian
is a single-carrier operator $\tilde{H}_{T}=\sum_{nn^{\prime}}\left(  \hat
{H}_{T}\right)  _{nn^{\prime}}a_{n}^{\dag}a_{n^{\prime}}$, the equation of
motion for $\left(  \hat{\rho}_{T}\right)  _{n^{\prime}n}$ reads $i\hbar
\frac{\partial}{\partial t}\left(  \hat{\rho}_{T}\right)  _{n^{\prime}%
n}=\left[  \hat{H}_{T},\hat{\rho}_{T}\right]  _{n^{\prime}n}$. The $n$
representation in the single-carrier Hilbert space is arbitrary thus
\begin{equation}
i\hbar\frac{\partial}{\partial t}\hat{\rho}_{T}=\left[  \hat{H}_{T},\hat{\rho
}_{T}\right]  \label{single-particle QLE}%
\end{equation}
with the operators acting on the single-carrier space. $\hat{\rho}_{T}$
satisfies $\left(  \hat{\rho}_{T}\right)  _{nn}=\left\langle N_{n}%
\right\rangle \geq0$ and $tr\hat{\rho}_{T}=N_{c}$ with $N_{n}=a_{n}^{\dag
}a_{n}$ and $\tilde{N}=\sum_{n}N_{n}$. Although normalized to the carrier
number $N_{c}$ instead of 1, $\hat{\rho}_{T}$ is often referred to as the
single-particle density matrix, the diagonal elements of which represent the
average occupation numbers of single-particle eigenstates rather than
occupation probability. This character implies that $\hat{\rho}_{T}$ is a
quantum-statistics generalization of the single-particle density function
described by the classical Boltzmann equation, and the diagonal elements of
$\hat{\rho}_{T}$ may comply with a Boltzmann-type transport equation. This
observation motivates one to split the quantum Liouville equation in the Bloch
representation into diagonal and off-diagonal parts in the following.

The single-carrier Hamiltonian reads $\hat{H}_{T}=\hat{H}_{0}+\hat{H}^{\prime
}+\hat{H}_{F}$, where $\hat{H}_{0}$ is the single-particle free Hamiltonian,
$\hat{H}^{\prime}=\lambda\hat{V}$ with $\lambda$ a dimensionless parameter and
$\hat{V}$ the disorder potential, and the field term $\hat{H}_{F}=\hat{H}%
_{1}e^{st}$ with $\hat{H}_{1}=-e\mathbf{E\cdot r}$ arises from the electric
field adiabatically switched-on from the remote past $t=-\infty$. The
infinitesimal positive $s$ in $\hat{H}_{F}$ can be taken to be the same as the
$s$ which appears as a regularization factor in the T-matrix theory of the
Boltzmann formalism \cite{Sinitsyn2007,Xiao2017AHE}. This is because the
physical situation is obtained by taking the limit $s\rightarrow0^{+}$. We
remind that a similar note on the infinitesimal positive $s$ has appeared in
the derivation of Kubo-Streda linear response formula with respect to the
uniform static electric field \cite{Streda2010}.

In the linear response regime one can thus decompose $\hat{\rho}_{T}$ into
\cite{KL1957}%
\begin{equation}
\hat{\rho}_{T}=\hat{\rho}+\hat{f}e^{st}, \label{density matrix}%
\end{equation}
where $\hat{\rho}$ is the equilibrium density matrix, $\hat{f}$ is the
out-of-equilibrium density matrix linear in the electric field at the time of
interest ($t=0$). The linear response of a single-particle observable $A$ thus
reads \cite{note-SOscattering}%
\begin{equation}
\delta A=tr\left\langle \hat{f}\hat{A}\right\rangle =\sum_{l}\left\langle
f_{l}\right\rangle A_{ll}+\sum_{ll^{\prime}}^{\prime}\left\langle
f_{ll^{\prime}}\right\rangle A_{l^{\prime}l}%
\end{equation}
in the eigenbasis of $\hat{H}_{0}$, where the index $l$ denotes the Bloch
state. Hereafter $\left\langle ..\right\rangle $ stands for disorder average,
and the notation $\sum^{\prime}$ means that all the index equalities should be
avoided in the summation. For anomalous and spin Hall conductivities in the
presence of weak Gaussian disorder, the leading contribution of $\delta A$ is
of $O\left(  \lambda^{0}\right)  $.

It is noticed that, $\rho_{ll^{\prime}}\neq\rho_{l}^{\left(  0\right)  }%
\delta_{ll^{\prime}}$ with $\rho_{l}^{\left(  0\right)  }$ the Fermi
distribution function, because $\rho_{ll^{\prime}}$ is altered by disorder and
thus even possesses off-diagonal elements. The neglect of this fact would lead
to the absence of an important interband-coherence process (Sec. II. C). In
fact, $\hat{\rho}$ can be expanded in the Bloch representation as:%
\begin{equation}
\rho_{ll^{\prime}}=\rho_{ll^{\prime}}^{\left(  0\right)  }+\rho_{ll^{\prime}%
}^{\left(  1\right)  }+\rho_{ll^{\prime}}^{\left(  2\right)  }+...,
\end{equation}
where the superscript means the order of $\lambda$. Here $\rho_{ll^{\prime}%
}^{\left(  0\right)  }=\rho_{l}^{\left(  0\right)  }\delta_{ll^{\prime}}$ is
known from the definition $\rho_{ll^{\prime}}^{\left(  0\right)  }=Tr\left(
a_{l^{\prime}}^{\dag}a_{l}\tilde{\rho}^{\left(  0\right)  }\right)  $.
Disorder-induced corrections $\rho_{ll^{\prime}}^{\left(  1\right)  ,\left(
2\right)  }$ can be obtained from an iterative solution to the quantum
Liouville equation $\left[  \hat{H}_{0}+\hat{H}^{\prime},\hat{\rho}\right]
=0$ obeyed by the equilibrium single-particle density matrix. The iteration
gives $\left[  \hat{H}_{0},\hat{\rho}^{\left(  1\right)  }\right]  =\left[
\hat{\rho}^{\left(  0\right)  },\hat{H}^{\prime}\right]  $ and $\left[
\hat{H}_{0},\hat{\rho}^{\left(  2\right)  }\right]  =\left[  \hat{\rho
}^{\left(  1\right)  },\hat{H}^{\prime}\right]  $, thus
\begin{equation}
\rho_{ll^{\prime}}^{\left(  1\right)  }=\frac{\rho_{l}^{\left(  0\right)
}-\rho_{l^{\prime}}^{\left(  0\right)  }}{d_{ll^{\prime}}}H_{ll^{\prime}%
}^{^{\prime}},\label{rho-1}%
\end{equation}
and $\rho_{ll^{\prime}}^{\left(  2\right)  }=\sum_{l^{\prime\prime}}%
^{^{\prime}}\frac{H_{ll^{\prime\prime}}^{^{\prime}}H_{l^{\prime\prime
}l^{\prime}}^{^{\prime}}}{d_{ll^{\prime}}}\left(  \frac{\rho_{l}%
-\rho_{l^{\prime\prime}}}{d_{ll^{\prime\prime}}}-\frac{\rho_{l^{\prime\prime}%
}-\rho_{l^{\prime}}}{d_{l^{\prime\prime}l^{\prime}}}\right)  $.\ Here
$\rho_{ll}^{\left(  1\right)  }=H_{ll}^{^{\prime}}\partial_{\epsilon_{l}}%
\rho_{l}^{\left(  0\right)  }=0$ ($H_{ll}^{^{\prime}}=0$, see below) has been
used, and $\rho_{ll}^{\left(  2\right)  }$ can be obtained from $\lim
_{l^{\prime}\rightarrow l}\rho_{ll^{\prime}}^{\left(  2\right)  }$. Hereafter
$d_{ll^{\prime}}\equiv\epsilon_{l}-\epsilon_{l^{\prime}}$, $d_{ll^{\prime}%
}^{\pm}\equiv d_{ll^{\prime}}\pm i\hbar s$ with $\epsilon_{l}$ the eigen-energy.

Now we turn to the out-of-equilibrium density matrix, which satisfies the
quantum Liouville equation
\begin{equation}
d_{ll^{\prime}}^{-}f_{ll^{\prime}}=\sum_{l^{\prime\prime}}\left(
f_{ll^{\prime\prime}}H_{l^{\prime\prime}l^{\prime}}^{\prime}-H_{ll^{\prime
\prime}}^{\prime}f_{l^{\prime\prime}l^{\prime}}\right)  +C_{ll^{\prime}}
\label{KL}%
\end{equation}
in the Bloch representation. $C_{ll^{\prime}}\equiv\left[  \hat{\rho},\hat
{H}_{1}\right]  _{ll^{\prime}}$ combines the electric field and equilibrium
density matrix, reading
\begin{equation}
C_{ll^{\prime}}=ie\mathbf{E}\cdot\left[  \left(  \partial_{\mathbf{k}%
}+\partial_{\mathbf{k}^{\prime}}\right)  \rho_{ll^{\prime}}+\left[
\mathbf{J},\hat{\rho}\right]  _{ll^{\prime}}\right]
\end{equation}
for $l\neq l^{\prime}$ and $C_{l}=ie\mathbf{E}\cdot\left[  \partial
_{\mathbf{k}}\rho_{ll}+\left[  \mathbf{J},\hat{\rho}\right]  _{ll}\right]  $,
where $\left[  \mathbf{J},\hat{\rho}\right]  _{ll^{\prime}}\equiv
\sum_{l^{\prime\prime}}\left(  \mathbf{J}_{ll^{\prime\prime}}\rho
_{l^{\prime\prime}l^{\prime}}-\rho_{ll^{\prime\prime}}\mathbf{J}%
_{l^{\prime\prime}l^{\prime}}\right)  $. Here $\mathbf{r}_{ll^{\prime}}%
=i\frac{\partial}{\partial\mathbf{k}}\delta_{ll^{\prime}}+i\mathbf{J}%
_{ll^{\prime}}$\ and $\mathbf{J}_{ll^{\prime}}\equiv\delta_{\mathbf{kk}%
^{\prime}}\langle u_{l}|\partial_{\mathbf{k}}|u_{l^{\prime}}\rangle$ are used.
$|l\rangle=|\mathbf{k}\rangle|u_{l}\rangle$ is the Bloch state, $l=\left(
\eta,\mathbf{k}\right)  $.\ Equation (\ref{KL}) can be split into
\cite{KL1957}
\begin{equation}
d_{ll^{\prime}}^{-}f_{ll^{\prime}}=\sum_{l^{\prime\prime}}^{\prime}\left(
f_{ll^{\prime\prime}}H_{l^{\prime\prime}l^{\prime}}^{\prime}-H_{ll^{\prime
\prime}}^{\prime}f_{l^{\prime\prime}l^{\prime}}\right)  +\left(
f_{l}-f_{l^{\prime}}\right)  H_{ll^{\prime}}^{\prime}+C_{ll^{\prime}}
\label{KL-offdiagonal}%
\end{equation}
for $l\neq l^{\prime}$, and
\begin{equation}
-i\hbar sf_{l}=\sum_{l^{\prime}}^{\prime}\left(  f_{ll^{\prime}}H_{l^{\prime
}l}^{\prime}-H_{ll^{\prime}}^{\prime}f_{l^{\prime}l}\right)  +C_{l}.
\label{KL-diagonal}%
\end{equation}
Here $H_{ll}^{\prime}$, which is the first-order energy correction in the bare
quantum mechanical perturbation theory, has been absorbed into $H_{0}$, thus
$H_{ll}^{\prime}=0$ hereafter.

In the case of weak disorder-potential, an iterative analysis of Eqs.
(\ref{KL-offdiagonal}) and (\ref{KL-diagonal}) in terms of the parameter
$\lambda$ is possible. First a starting point for this iteration is needed. To
do this one has to assume that, the most conventional Boltzmann equation
(where the scattering amplitude is obtained under the lowest-order Born
approximation) gives the leading order contribution to $f_{l}$, i.e., $f_{l}%
$\ starts from the order of $\lambda^{-2}$. In other words, we demand the most
conventional Boltzmann equation is at least qualitatively correct as a leading
approximation of longitudinal electronic transport. This is always true for
the electrical conductivity in the metallic regime where the Fermi energy is
much larger than the disorder-induced band broadening. However, this
assumption may break down for spin-orbit torques when there are multiple
spin-orbit-split bands on the Fermi surface if the minimal interband splitting
is smaller than the disorder-induced band broadening (weak spin-orbit-coupling
regime), even in the metallic regime \cite{Xiao2017SOT}. In that case the
field-like torque may not be captured by the conventional Boltzmann equation
(detailed discussions in Ref. \cite{Xiao2017SOT}). Therefore, a necessary
condition for the validity of the Boltzmann formalism is that the minimal
interband splitting around the Fermi level is smaller than the
disorder-induced band broadening.

Then $sf_{l}\rightarrow0$ when $s\rightarrow0^{+}$, if the electric field is
turned on much more slowly than the scattering time \cite{KL1957,Moore1967}.
And $f_{ll^{\prime}}$\ starts from the order of $\lambda^{-1}$.\ Thus an
order-by-order analysis with respect to the weak disorder potential follows:%
\begin{align}
f_{l}  &  =f_{l}^{\left(  -2\right)  }+f_{l}^{\left(  -1\right)  }%
+f_{l}^{\left(  0\right)  }+...,\nonumber\\
f_{ll^{\prime}}  &  =f_{ll^{\prime}}^{\left(  -1\right)  }+f_{ll^{\prime}%
}^{\left(  0\right)  }+f_{ll^{\prime}}^{\left(  1\right)  }...\text{\ }\left(
l\neq l^{\prime}\right)  ,\\
C_{ll^{\prime}}  &  =C_{ll^{\prime}}^{\left(  0\right)  }+C_{ll^{\prime}%
}^{\left(  1\right)  }+C_{ll^{\prime}}^{\left(  2\right)  }+...\text{
\ }.\nonumber
\end{align}
The iteration yields an equation only concerning the diagonal element $f_{l}$
and a series of equations expressing $f_{ll^{\prime}}$ in terms of $f_{l}$
\cite{Luttinger1958,Sinitsyn2008}, e.g., $f_{ll^{\prime}}^{\left(  -1\right)
}=\frac{f_{l}^{\left(  -2\right)  }-f_{l^{\prime}}^{\left(  -2\right)  }%
}{d_{ll^{\prime}}^{-}}H_{ll^{\prime}}^{\prime}$, and%
\begin{align}
f_{ll^{\prime}}^{\left(  0\right)  }  &  =\sum_{l^{\prime\prime}}^{\prime
}\frac{H_{ll^{\prime\prime}}^{\prime}H_{l^{\prime\prime}l^{\prime}}^{\prime}%
}{d_{ll^{\prime}}^{-}}\left[  \frac{f_{l}^{\left(  -2\right)  }-f_{l^{\prime
\prime}}^{\left(  -2\right)  }}{d_{ll^{\prime\prime}}^{-}}-\frac
{f_{l^{\prime\prime}}^{\left(  -2\right)  }-f_{l^{\prime}}^{\left(  -2\right)
}}{d_{l^{\prime\prime}l^{\prime}}^{-}}\right] \nonumber\\
&  +\frac{f_{l}^{\left(  -1\right)  }-f_{l^{\prime}}^{\left(  -1\right)  }%
}{d_{ll^{\prime}}^{-}}H_{ll^{\prime}}^{\prime}+\frac{C_{ll^{\prime}}^{\left(
0\right)  }}{d_{ll^{\prime}}^{-}}. \label{off0}%
\end{align}

The required transport equation for $\left\langle f_{l}\right\rangle $ is
obtained by disorder averaging. In so doing, one assumes that $f_{l}$ does not
contain any physically important, rapidly varying exponential factors, thus in
the thermodynamic limit $\left\langle f_{l}^{\left(  -2\right)  }%
H_{ll^{\prime}}^{\prime}\right\rangle =\left\langle f_{l}^{\left(  -2\right)
}\right\rangle \left\langle H_{ll^{\prime}}^{\prime}\right\rangle $,
$\left\langle H_{ll^{\prime\prime}}^{\prime}H_{l^{\prime\prime}l^{\prime}%
}^{\prime}f_{l}^{\left(  -2\right)  }\right\rangle =\left\langle
H_{ll^{\prime\prime}}^{\prime}H_{l^{\prime\prime}l^{\prime}}^{\prime
}\right\rangle \left\langle f_{l}^{\left(  -2\right)  }\right\rangle $. Only
when this assumption is true, the semiclassical distribution function and thus
the Boltzmann formalism can be defined. The validity of this assumption has
been confirmed \cite{Moore1967}, but beyond the scope of our study. Up to
$\left\langle f_{l}^{\left(  0\right)  }\right\rangle $ we get
\begin{align*}
0  &  =\frac{1}{i\hbar}C_{l}^{\left(  0\right)  }+\sum_{l^{\prime}}\left[
\omega_{l^{\prime}l}^{\left(  2\right)  }\left\langle f_{l}^{\left(
-2\right)  }\right\rangle -\omega_{ll^{\prime}}^{\left(  2\right)
}\left\langle f_{l^{\prime}}^{\left(  -2\right)  }\right\rangle \right] \\
&  +\sum_{l^{\prime}}\left\{  \left[  \omega_{l^{\prime}l}^{\left(  4\right)
}+S_{l^{\prime}l}^{\left(  4\right)  }\right]  \left\langle f_{l}^{\left(
-2\right)  }\right\rangle -\left[  \omega_{ll^{\prime}}^{\left(  4\right)
}+S_{ll^{\prime}}^{\left(  4\right)  }\right]  \left\langle f_{l^{\prime}%
}^{\left(  -2\right)  }\right\rangle \right\} \\
&  +\frac{1}{i\hbar}C_{l}^{\prime\prime}+\sum_{l^{\prime}}\left[
\omega_{l^{\prime}l}^{\left(  2\right)  }\left\langle f_{l}^{\left(  0\right)
}\right\rangle -\omega_{ll^{\prime}}^{\left(  2\right)  }\left\langle
f_{l^{\prime}}^{\left(  0\right)  }\right\rangle \right]  ,
\end{align*}
where $C_{l}^{\left(  0\right)  }=ie\mathbf{E}\cdot\partial_{\mathbf{k}}%
\rho_{l}^{\left(  0\right)  }$ is the conventional driving term of the
Boltzmann equation, and $\omega_{ll^{\prime}}^{\left(  2\right)  }%
=\omega_{ll^{\prime}}^{\left(  2\right)  }=\frac{2\pi}{\hbar}\left\langle
\left\vert H_{ll^{\prime}}^{\prime}\right\vert ^{2}\right\rangle \delta\left(
d_{ll^{\prime}}\right)  $. The anomalous driving term $C_{l}^{\prime\prime}$
contains combination effects of the electric field and disorder, and is of
$O\left(  \lambda^{2}\right)  $ \cite{Luttinger1958}:
\begin{align}
C_{l}^{^{\prime\prime}}  &  \equiv\sum_{l^{\prime\prime}l^{\prime}}^{^{\prime
}}\left\{  \left[  \frac{\left\langle C_{ll^{\prime\prime}}^{\left(  0\right)
}H_{l^{\prime\prime}l^{\prime}}^{^{\prime}}H_{l^{\prime}l}^{^{\prime}%
}\right\rangle }{d_{ll^{\prime\prime}}^{-}d_{ll^{\prime}}^{-}}-\frac
{\left\langle H_{ll^{\prime\prime}}^{^{\prime}}C_{l^{\prime\prime}l^{\prime}%
}^{\left(  0\right)  }H_{l^{\prime}l}^{^{\prime}}\right\rangle }%
{d_{l^{\prime\prime}l^{\prime}}^{-}d_{ll^{\prime}}^{-}}\right]  \right.
\nonumber\\
&  \left.  -c.c\right\}  +\sum_{l^{\prime}}^{^{\prime}}\left[  \frac
{\left\langle C_{ll^{\prime}}^{\left(  1\right)  }H_{l^{\prime}l}^{^{\prime}%
}\right\rangle }{d_{ll^{\prime}}^{-}}-c.c.\right]  +\left\langle
C_{l}^{\left(  2\right)  }\right\rangle ,
\end{align}
where%
\begin{align*}
C_{ll^{\prime}}^{\left(  0\right)  }  &  =ie\mathbf{E\cdot J}_{ll^{\prime}%
}\left(  \rho_{l^{\prime}}^{\left(  0\right)  }-\rho_{l}^{\left(  0\right)
}\right)  ,\\
C_{ll^{\prime}}^{\left(  1\right)  }  &  =ie\mathbf{E\cdot}\left[
\mathbf{\hat{D}}\rho_{ll^{\prime}}^{\left(  1\right)  }+\sum_{l^{\prime\prime
}}\left(  \mathbf{J}_{ll^{\prime\prime}}\rho_{l^{\prime\prime}l^{\prime}%
}^{\left(  1\right)  }-\rho_{ll^{\prime\prime}}^{\left(  1\right)  }%
\mathbf{J}_{l^{\prime\prime}l^{\prime}}\right)  \right]  ,\\
C_{l}^{\left(  2\right)  }  &  =ie\mathbf{E\cdot}\left[  \partial_{\mathbf{k}%
}\rho_{ll}^{\left(  2\right)  }+\sum_{l^{\prime}}\left(  \mathbf{J}%
_{ll^{\prime}}\rho_{l^{\prime}l}^{\left(  2\right)  }-\rho_{ll^{\prime}%
}^{\left(  2\right)  }\mathbf{J}_{l^{\prime}l}\right)  \right]  ,
\end{align*}
with $\mathbf{\hat{D}}=\partial_{\mathbf{k}}+\partial_{\mathbf{k}^{\prime}}$.
After some delicate steps we arrive at Luttinger's expression (see Appendix)
for $C_{l}^{\prime\prime}$. Then we throw away all the terms which still exist
in the case of non-relativistic free electrons. These terms are trivial
renormalizations to the conventional driving term, which only contribute to
the spin-orbit induced Hall transport in $O\left(  \lambda^{2}\right)  $.
Hence%
\begin{gather}
\frac{1}{i\hbar}C_{l}^{\prime\prime}=e\mathbf{E\cdot}\sum_{l^{\prime}}%
\frac{2\pi}{\hbar}\left\langle \left\vert H_{ll^{\prime}}^{\prime}\right\vert
^{2}\right\rangle \delta\left(  \epsilon_{l}-\epsilon_{l^{\prime}}\right)
\nonumber\\
\times\left(  i\mathbf{J}_{l^{\prime}}-i\mathbf{J}_{l}-\mathbf{\hat{D}}\arg
H_{l^{\prime}l}^{\prime}\right)  \left(  -\frac{\partial\rho_{l}^{\left(
0\right)  }}{\partial\epsilon_{l}}\right)  .
\end{gather}
One can recognize that the quantity
\begin{equation}
\delta\mathbf{r}_{l^{\prime}l}=i\mathbf{J}_{l^{\prime}}-i\mathbf{J}%
_{l}-\mathbf{\hat{D}}\arg H_{l^{\prime}l}^{^{\prime}}%
\end{equation}
has a definite semiclassical meaning as the coordinate-shift of the
center-of-mass of electron wavepackets during the scattering
\cite{Sinitsyn2006,Ivchenko1982,Geller1988}. Thus
\begin{equation}
C_{l}^{\prime\prime}=-i\hbar e\mathbf{E\cdot v}_{l}^{sj}\partial_{\epsilon
_{l}}\rho_{l}^{\left(  0\right)  }, \label{anomalous drive}%
\end{equation}
where $\mathbf{v}_{l}^{sj}=\sum_{l^{\prime}}^{\prime}\omega_{ll^{\prime}%
}^{\left(  2\right)  }\delta\mathbf{r}_{l^{\prime}l}$ just coincides with the
semiclassically constructed side-jump velocity in the Boltzmann theory of the
anomalous Hall effect \cite{Sinitsyn2006}. The interband-coherence nature of
the anomalous driving term $C_{l}^{\prime\prime}$ can be seen when
$\mathbf{v}_{l}^{sj}$ is identified in a more general form in the following subsection.

Some very delicate algebra leads to Eq. (\ref{scattering-rate-4}) for
$\omega_{l^{\prime}l}^{\left(  4\right)  }$ and $\omega_{ll^{\prime}}^{\left(
4\right)  }$. The expression for $S_{ll^{\prime}}^{\left(  4\right)
}=S_{l^{\prime}l}^{\left(  4\right)  }$ \cite{KL1957,Luttinger1958} is not
necessary here. We note that $\omega_{ll^{\prime}}$ coincides with the
semiclassical scattering rate in the case of Gaussian disorder calculated by
the golden rule \cite{Sinitsyn2008,Xiao2017AHE}
\begin{equation}
\omega_{ll^{\prime}}=\frac{2\pi}{\hbar}\left\langle \left\vert T_{ll^{\prime}%
}\right\vert ^{2}\right\rangle \delta\left(  d_{ll^{\prime}}\right)
=\omega_{ll^{\prime}}^{\left(  2\right)  }+\omega_{ll^{\prime}}^{\left(
4\right)  }+..., \label{golden-rule}%
\end{equation}
where $T_{ll^{\prime}}$ is the T-matrix in the single-particle scattering
theory. The symmetric parts ($\omega_{l^{\prime}l}^{s}=\frac{1}{2}\left(
\omega_{l^{\prime}l}+\omega_{ll^{\prime}}\right)  $ and $S_{ll^{\prime}%
}^{\left(  4\right)  }$) of the higher-order scattering rates ($\omega
_{l^{\prime}l}^{\left(  4\right)  }$, $\omega_{ll^{\prime}}^{\left(  4\right)
}$ and $S_{ll^{\prime}}^{\left(  4\right)  }$) only contribute to the
spin-orbit induced Hall transport in $O\left(  \lambda^{2}\right)  $, and thus
can be neglected in the case of weak disorder potential. In contrast, the
anti-symmetric parts ($\omega_{l^{\prime}l}^{a}=\frac{1}{2}\left(
\omega_{l^{\prime}l}-\omega_{ll^{\prime}}\right)  $) of $\omega_{l^{\prime}%
l}^{\left(  4\right)  }$ and $\omega_{ll^{\prime}}^{\left(  4\right)  }$
contribute to the spin-orbit induced Hall transport in $O\left(  \lambda
^{0}\right)  $, the leading order in the case of Gaussian disorder. Assuming
isotropic systems $\sum_{l^{\prime}}\omega_{l^{\prime}l}^{a}=0$, one has%
\begin{gather}
-e\mathbf{E}\cdot\mathbf{v}_{l}\partial_{\epsilon_{l}}\rho_{l}^{\left(
0\right)  }=\sum_{l^{\prime}}\omega_{ll^{\prime}}^{\left(  2\right)  }\left[
\left\langle f_{l}^{\left(  -2\right)  }\right\rangle -\left\langle
f_{l^{\prime}}^{\left(  -2\right)  }\right\rangle \right]  ,\nonumber\\
\sum_{l^{\prime}}\omega_{ll^{\prime}}^{\left(  2\right)  }\left\langle
f_{l}^{\left(  0\right)  ,n}-f_{l^{\prime}}^{\left(  0\right)  ,n}%
\right\rangle =-\sum_{l^{\prime}}\omega_{ll^{\prime}}^{4a}\left\langle
f_{l}^{\left(  -2\right)  }-f_{l^{\prime}}^{\left(  -2\right)  }\right\rangle
,\nonumber\\
e\mathbf{E}\cdot\mathbf{v}_{l}^{sj}\partial_{\epsilon_{l}}\rho_{l}^{\left(
0\right)  }=\sum_{l^{\prime}}\omega_{ll^{\prime}}^{\left(  2\right)
}\left\langle f_{l}^{\left(  0\right)  ,a}-f_{l^{\prime}}^{\left(  0\right)
,a}\right\rangle
\end{gather}
for%
\begin{equation}
\left\langle f_{l}\right\rangle =\left\langle f_{l}^{\left(  -2\right)
}\right\rangle +\left\langle f_{l}^{\left(  0\right)  ,n}\right\rangle
+\left\langle f_{l}^{\left(  0\right)  ,a}\right\rangle \label{diagonal f}%
\end{equation}
in the weak disorder-potential regime. Here $f_{l}^{\left(  0\right)  }%
=f_{l}^{\left(  0\right)  ,n}+f_{l}^{\left(  0\right)  ,a}$,
where$\ \left\langle f_{l}^{\left(  0\right)  ,n}\right\rangle $ arises from
the anti-symmetric part of the differential scattering cross-section
$\omega_{ll^{\prime}}^{4a}=\frac{1}{2}\left[  \omega_{ll^{\prime}}^{\left(
4\right)  }-\omega_{l^{\prime}l}^{\left(  4\right)  }\right]  $,
and$\ \left\langle f_{l}^{\left(  0\right)  ,a}\right\rangle $ describes the
response to the coordinate-shift induced anomalous driving term. These three
equations just constitute the modified Boltzmann equation
\cite{Sinitsyn2008,Nagaosa2010}.

\subsection{Interband mixing of Bloch states and interband elements of the
out-of-equilibrium density matrix}

Equation (\ref{off0}) yields ($\left\langle f_{ll^{\prime}}^{\left(
-1\right)  }\right\rangle =0$)%
\begin{gather*}
\sum_{ll^{\prime}}^{\prime}\left\langle f_{ll^{\prime}}\right\rangle
A_{l^{\prime}l}=\sum_{ll^{\prime}}^{\prime}\left\langle f_{ll^{\prime}%
}^{\left(  0\right)  }\right\rangle A_{l^{\prime}l}=\sum_{ll^{\prime}}%
^{\prime}C_{ll^{\prime}}^{\left(  0\right)  }\frac{A_{l^{\prime}l}%
}{d_{ll^{\prime}}^{-}}+\\
\sum_{ll^{\prime}l^{\prime\prime}}^{\prime}\left\langle \frac{f_{l}^{\left(
-2\right)  }-f_{l^{\prime\prime}}^{\left(  -2\right)  }}{d_{ll^{\prime\prime}%
}^{-}}-\frac{f_{l^{\prime\prime}}^{\left(  -2\right)  }-f_{l^{\prime}%
}^{\left(  -2\right)  }}{d_{l^{\prime\prime}l^{\prime}}^{-}}\right\rangle
\frac{\left\langle H_{ll^{\prime\prime}}^{\prime}H_{l^{\prime\prime}l^{\prime
}}^{\prime}\right\rangle A_{l^{\prime}l}}{d_{ll^{\prime}}^{-}},
\end{gather*}
respectively. Due to $\mathbf{v}_{ll^{\prime}}\delta_{\mathbf{kk}^{\prime}%
}=-\frac{1}{\hbar}d_{ll^{\prime}}\mathbf{J}_{ll^{\prime}}$ for $l\neq
l^{\prime}$, we have $\sum_{ll^{\prime}}^{\prime}C_{ll^{\prime}}^{\left(
0\right)  }A_{l^{\prime}l}/d_{ll^{\prime}}^{-}=\sum_{l}\rho_{l}^{\left(
0\right)  }\delta^{in}A_{l}$, where
\begin{equation}
\delta^{in}A_{l}=-\hbar e\mathbf{E\cdot}\sum_{l^{\prime}\neq l}2\mathrm{Im}%
\langle u_{l}|\mathbf{\hat{v}}|u_{l^{\prime}}\rangle\delta_{\mathbf{kk}%
^{\prime}}A_{l^{\prime}l}/d_{ll^{\prime}}^{2} \label{in}%
\end{equation}
coincides with the correction to $A_{l}$ due to electric-field-induced
interband mixing of Bloch states introduced in the Boltzmann theory
\cite{Lee2015,Xiao2017SHE,Xiao2017SOT-SBE}. Besides, by interchanging the
indices $l$, $l^{\prime}$ and $l^{\prime\prime}$ here and there and some
simple algebra, we find%
\begin{align}
&  \sum_{ll^{\prime}l^{\prime\prime}}^{\prime}\left\langle H_{ll^{\prime
\prime}}^{\prime}H_{l^{\prime\prime}l^{\prime}}^{\prime}\right\rangle
\left\langle \frac{f_{l}^{\left(  -2\right)  }-f_{l^{\prime\prime}}^{\left(
-2\right)  }}{d_{ll^{\prime\prime}}^{-}}-\frac{f_{l^{\prime\prime}}^{\left(
-2\right)  }-f_{l^{\prime}}^{\left(  -2\right)  }}{d_{l^{\prime\prime
}l^{\prime}}^{-}}\right\rangle \frac{A_{l^{\prime}l}}{d_{ll^{\prime}}^{-}%
}\nonumber\\
&  =\sum_{ll^{\prime}l^{\prime\prime}}^{\prime}\left\langle f_{l}^{\left(
-2\right)  }\right\rangle \left[  \frac{\left\langle H_{l^{\prime}%
l^{\prime\prime}}^{\prime}H_{l^{\prime\prime}l}^{\prime}\right\rangle
A_{ll^{\prime}}}{d_{ll^{\prime}}^{+}d_{ll^{\prime\prime}}^{+}}+c.c.\right]
\nonumber\\
&  +\sum_{ll^{\prime}l^{\prime\prime}}^{\prime}\left\langle f_{l}^{\left(
-2\right)  }\right\rangle \left\langle H_{l^{\prime\prime}l}^{\prime
}H_{ll^{\prime}}^{\prime}\right\rangle A_{l^{\prime}l^{\prime\prime}}\left(
\frac{1}{d_{ll^{\prime\prime}}^{+}}-\frac{1}{d_{ll^{\prime}}^{-}}\right)
\frac{1}{d_{l^{\prime\prime}l^{\prime}}^{-}}\nonumber\\
&  =\sum_{l}\left\langle f_{l}^{\left(  -2\right)  }\right\rangle \delta
^{ex}A_{l},
\end{align}
where
\begin{equation}
\delta^{ex}A_{l}=\sum_{l^{\prime}l^{\prime\prime}}^{\prime}\left[
2\operatorname{Re}\frac{\left\langle H_{l^{\prime}l^{\prime\prime}}^{\prime
}H_{l^{\prime\prime}l}^{\prime}\right\rangle A_{ll^{\prime}}}{d_{ll^{\prime}%
}^{+}d_{ll^{\prime\prime}}^{+}}+\frac{\left\langle H_{l^{\prime\prime}%
l}^{\prime}H_{ll^{\prime}}^{\prime}\right\rangle A_{l^{\prime}l^{\prime\prime
}}}{d_{ll^{\prime\prime}}^{+}d_{ll^{\prime}}^{-}}\right]  \label{ex}%
\end{equation}
coincides with the correction to $A_{l}$ due to disorder-induced interband
mixing of Bloch states introduced semi-phenomenologically in the Boltzmann
theory \cite{Xiao2017SHE,Xiao2017SOT-SBE}. Therefore, in the weak
disorder-potential regime
\begin{equation}
\sum_{ll^{\prime}}^{\prime}\left\langle f_{ll^{\prime}}\right\rangle
A_{l^{\prime}l}=\sum_{l}\rho_{l}^{\left(  0\right)  }\delta^{in}A_{l}+\sum
_{l}\left\langle f_{l}^{\left(  -2\right)  }\right\rangle \delta^{ex}A_{l}.
\label{Culcer}%
\end{equation}
In the anomalous Hall effect ($\hat{A}=\mathbf{\hat{v}}$), $\delta
^{in}\mathbf{v}_{l}$ and $\delta^{ex}\mathbf{v}_{l}$ are just the
Berry-curvature anomalous velocity and side-jump velocity
\cite{Sinitsyn2006,Xiao2017SOT-SBE}, respectively. Thus the anomalous driving
term (\ref{anomalous drive}) is just $e\mathbf{E}\cdot\delta^{ex}%
\mathbf{v}_{l}\partial_{\epsilon_{l}}\rho_{l}^{\left(  0\right)  }$, of which
the interband-coherence nature is apparent.

We also note that in a recent alternative density matrix treatment to
interband-coherence responses \cite{Culcer2017}, where only interband elements
of the \textit{out-of-equilibrium} density-matrix are considered, their main
results (Eqs. (45), (47) and (48) in Ref. \cite{Culcer2017}) just correspond
to our Eq. (\ref{Culcer}).

\subsection{Anomalous driving term and off-diagonal elements of the
equilibrium density matrix}

Now we look into the anomalous driving term $C_{l}^{\prime\prime}$.

$C_{ll^{^{\prime}}}^{\left(  0\right)  }$, $C_{ll^{^{\prime}}}^{\left(
1\right)  }$ and $C_{l}^{\left(  2\right)  }$ contain both the diagonal
($\mathbf{\hat{D}}$, $\partial_{\mathbf{k}}$\ and $\mathbf{J}_{ll}$) and
off-diagonal ($\mathbf{J}_{ll^{\prime}}$) components of the electric-field
perturbation $\hat{H}_{1}$ in the Bloch representation, as well as diagonal
and off-diagonal components of the equilibrium density matrix. We notice that,
if we demand that only the diagonal (also band-diagonal) components of the
electric-field perturbation contribute to the final form of $C_{l}%
^{\prime\prime}$, i.e., we only preserve $C_{ll^{^{\prime}}}^{\left(
0\right)  }\sim0$, $C_{l}^{\left(  2\right)  }\sim ie\mathbf{E}\cdot
\partial_{\mathbf{k}}\rho_{ll}^{\left(  2\right)  }$ and $C_{ll^{\prime}%
}^{\left(  1\right)  }\sim ie\mathbf{E\cdot}\left[  \mathbf{\hat{D}}%
\rho_{ll^{\prime}}^{\left(  1\right)  }+\left(  \mathbf{J}_{l}-\mathbf{J}%
_{l^{\prime}}\right)  \rho_{ll^{\prime}}^{\left(  1\right)  }\right]  $ (here
$l\neq l^{\prime}$), then we directly arrive at Luttinger's expression for
$C_{l}^{\prime\prime}$. If we further demand that only the off-diagonal
elements of the equilibrium density matrix survive in the final form of
$C_{l}^{\prime\prime}$, thus $C_{l}^{\left(  2\right)  }\sim0$ and
\begin{equation}
C_{l}^{^{\prime\prime}}=\sum_{l^{\prime}}^{^{\prime}}\left[  \left\langle
C_{ll^{\prime}}^{\left(  1\right)  }H_{l^{\prime}l}^{^{\prime}}\right\rangle
/d_{ll^{\prime}}^{-}-c.c.\right]
\end{equation}
with
\begin{equation}
C_{ll^{\prime}}^{\left(  1\right)  }=ie\mathbf{E\cdot}\left[  \mathbf{\hat{D}%
}\rho_{ll^{\prime}}^{\left(  1\right)  }+\left(  \mathbf{J}_{l}-\mathbf{J}%
_{l^{\prime}}\right)  \rho_{ll^{\prime}}^{\left(  1\right)  }\right]  ,
\end{equation}
and throw away the trivial renormalization terms, we again obtain Eq.
(\ref{anomalous drive}). Therefore, the anomalous driving term
(\ref{anomalous drive}) adopted in the modified Boltzmann formalism results in
fact from the combination of off-diagonal elements of the equilibrium density
matrix and the diagonal electric-field perturbations.

\subsection{Interband and intraband virtual scattering contributions to
intrinsic-skew-scattering}

In previous Boltzmann theories \cite{Sinitsyn2008,Xiao2017SOT-SBE}, the
anti-symmetric part $\omega_{ll^{\prime}}^{4a}$ of $\omega_{ll^{\prime}%
}^{\left(  4\right)  }$ was calculated within the noncrossing approximation
\cite{Xiao2017AHE,Sinitsyn2007}. Here we show that the crossed part of
$\omega_{ll^{\prime}}^{4a}$ corresponds to the recently identified Hall
contribution of X and $\Psi$ diagrams in the Kubo diagrammatic approach
\cite{Ado2015,Ado2016}.

Starting from Eq. (\ref{scattering-rate-4}) we have (set $\lambda=1$
hereafter)%
\begin{align*}
\omega_{ll^{\prime}}^{4a}  &  =-\frac{2\pi}{\hbar}\delta\left(  d_{ll^{\prime
}}\right)  \sum_{l^{\prime\prime},l^{\prime\prime\prime}}^{\prime}\left[
\mathrm{Im}\left\langle V_{ll^{\prime\prime\prime}}V_{l^{\prime\prime\prime
}l^{\prime}}V_{l^{\prime}l^{\prime\prime}}V_{l^{\prime\prime}l}\right\rangle
\mathrm{Im}\frac{1}{d_{ll^{\prime\prime}}^{-}d_{ll^{\prime\prime\prime}}^{+}%
}\right. \\
&  \left.  +\mathrm{Im}\left\langle V_{ll^{\prime}}V_{l^{\prime}%
l^{\prime\prime}}V_{l^{\prime\prime}l^{\prime\prime\prime}}V_{l^{\prime
\prime\prime}l}\right\rangle \mathrm{Im}\frac{1}{d_{ll^{\prime\prime}}%
^{-}d_{ll^{\prime\prime\prime}}^{-}}\right. \\
&  \left.  +\mathrm{Im}\left\langle V_{ll^{\prime\prime\prime}}V_{l^{\prime
\prime\prime}l^{\prime\prime}}V_{l^{\prime\prime}l^{\prime}}V_{l^{\prime}%
l}\right\rangle \mathrm{Im}\frac{1}{d_{ll^{\prime\prime}}^{+}d_{ll^{\prime
\prime\prime}}^{+}}\right]  .
\end{align*}
When taking the disorder average in the case of Gaussian disorder, there exist
both the non-crossed (nc) and crossed (c) contributions
\cite{Sinitsyn2007,Ado2015}, thus $\omega_{ll^{\prime}}^{4a}=\omega
_{ll^{\prime}}^{4a-nc}+\omega_{ll^{\prime}}^{4a-c}$. For example,
$\left\langle V_{ll^{\prime\prime\prime}}V_{l^{\prime\prime\prime}l^{\prime}%
}V_{l^{\prime}l^{\prime\prime}}V_{l^{\prime\prime}l}\right\rangle $ contains
noncrossed contribution \cite{Sinitsyn2007,Xiao2017AHE} $\left\langle
V_{ll^{\prime\prime\prime}}V_{l^{\prime\prime}l}\right\rangle \left\langle
V_{l^{\prime\prime\prime}l^{\prime}}V_{l^{\prime}l^{\prime\prime}%
}\right\rangle $ and crossed contribution $\left\langle V_{ll^{\prime
\prime\prime}}V_{l^{\prime}l^{\prime\prime}}\right\rangle \left\langle
V_{l^{\prime\prime\prime}l^{\prime}}V_{l^{\prime\prime}l}\right\rangle $
corresponding to the so-called X diagram \cite{Ado2016}, while $\left\langle
V_{ll^{\prime\prime\prime}}V_{l^{\prime\prime\prime}l^{\prime}}\right\rangle
\left\langle V_{l^{\prime}l^{\prime\prime}}V_{l^{\prime\prime}l}\right\rangle
\delta\left(  d_{ll^{\prime}}\right)  $ implies $l=l^{\prime}$ and thus does
not contribute to $\omega_{ll^{\prime}}^{4a}$. $\left\langle V_{ll^{\prime}%
}V_{l^{\prime}l^{\prime\prime}}V_{l^{\prime\prime}l^{\prime\prime\prime}%
}V_{l^{\prime\prime\prime}l}\right\rangle $ contains noncrossed contributions
\cite{Sinitsyn2007,Xiao2017AHE} $\left\langle V_{ll^{\prime}}V_{l^{\prime
}l^{\prime\prime}}\right\rangle \left\langle V_{l^{\prime\prime}%
l^{\prime\prime\prime}}V_{l^{\prime\prime\prime}l}\right\rangle $ and
$\left\langle V_{ll^{\prime}}V_{l^{\prime\prime\prime}l}\right\rangle
\left\langle V_{l^{\prime}l^{\prime\prime}}V_{l^{\prime\prime}l^{\prime
\prime\prime}}\right\rangle $ as well as crossed contribution $\left\langle
V_{ll^{\prime}}V_{l^{\prime\prime}l^{\prime\prime\prime}}\right\rangle
\left\langle V_{l^{\prime}l^{\prime\prime}}V_{l^{\prime\prime\prime}%
l}\right\rangle $ related to one $\Psi$ diagram \cite{Ado2016}.

Alternatively, we have
\begin{align*}
\omega_{ll^{\prime}}^{4a}  &  =-\frac{\left(  2\pi\right)  ^{2}}{\hbar}%
\delta\left(  d_{ll^{\prime}}\right)  \sum_{l^{\prime\prime},l^{\prime
\prime\prime}}^{\prime}\frac{\delta\left(  d_{ll^{\prime\prime}}\right)
}{d_{ll^{\prime\prime\prime}}}\left[  \mathrm{Im}\left\langle V_{ll^{\prime
\prime\prime}}V_{l^{\prime\prime\prime}l^{\prime}}V_{l^{\prime}l^{\prime
\prime}}V_{l^{\prime\prime}l}\right\rangle \right. \\
&  \left.  +\mathrm{Im}\left\langle V_{ll^{\prime}}V_{l^{\prime}%
l^{\prime\prime}}V_{l^{\prime\prime}l^{\prime\prime\prime}}V_{l^{\prime
\prime\prime}l}\right\rangle +\mathrm{Im}\left\langle V_{ll^{\prime}%
}V_{l^{\prime}l^{\prime\prime\prime}}V_{l^{\prime\prime\prime}l^{\prime\prime
}}V_{l^{\prime\prime}l}\right\rangle \right]  ,
\end{align*}
from which we see that the $l,l^{\prime}$ and $l^{\prime\prime}$ states lie on
the mass shell, while the $l^{\prime\prime\prime}$ state does not
\cite{Ado2017}. This means that virtual off-shell scattering are indispensable
for intrinsic-skew-scattering in the presence of Gaussian disorder. This
observation was first made in analyzing the Feynman diagrams
\cite{Ado2015,Ado2017,Konig2016}.

To be definite, we consider slowly varying scalar impurity potentials
\cite{Luttinger1958}, then%
\[
V_{ll^{\prime}}=V_{\mathbf{kk}^{\prime}}\left[  \delta_{\eta\eta^{\prime}%
}+\delta k_{\mu}J_{\mu}^{\eta\eta^{\prime}}\left(  \mathbf{k}\right)
+\frac{1}{2}\delta k_{\mu}\delta k_{\nu}J_{\mu\nu}^{\eta\eta^{\prime}}\left(
\mathbf{k}\right)  +...\right]  ,
\]
where $\delta k_{\mu}=k_{\mu}^{\prime}-k_{\mu}$, $J_{\mu}^{\eta\eta^{\prime}%
}\left(  \mathbf{k}\right)  =\langle u_{\eta\mathbf{k}}|\partial_{k_{\mu}%
}|u_{\eta^{\prime}\mathbf{k}}\rangle$ and $J_{\mu\nu}^{\eta\eta^{\prime}%
}\left(  \mathbf{k}\right)  =\langle u_{\eta\mathbf{k}}|\partial_{k_{\nu}%
}\partial_{k_{\mu}}|u_{\eta^{\prime}\mathbf{k}}\rangle$. The Einstein
summation convention is used hereafter for the indices $\mu$, $\nu$. The
noncrossed part contributes%
\begin{align}
\omega_{ll^{\prime}}^{4a-nc}  &  =\frac{\left(  2\pi n_{im}V_{0}^{2}\right)
^{2}}{2\hbar}\left(  \mathbf{k\times k}^{\prime}\right)  _{\mu\nu}\delta
_{\eta^{\prime}\eta}\delta\left(  d_{ll^{\prime}}\right) \nonumber\\
&  \times\sum_{l^{\prime\prime},l^{\prime\prime\prime}}^{\prime}\frac
{\delta_{\eta^{\prime\prime}\eta}\delta\left(  d_{ll^{\prime\prime}}\right)
}{d_{ll^{\prime\prime\prime}}}\left(  \delta_{\mathbf{k}^{\prime\prime
}\mathbf{k}^{\prime\prime\prime}}+\delta_{\mathbf{k}^{\prime}\mathbf{k}%
^{\prime\prime\prime}}+\delta_{\mathbf{kk}^{\prime\prime\prime}}\right)
\nonumber\\
&  \times\mathrm{Im}\left[  J_{\mu}^{\eta\eta^{\prime\prime\prime}}\left(
\mathbf{k}\right)  J_{\nu}^{\eta^{\prime\prime\prime}\eta}\left(
\mathbf{k}\right)  \right]  , \label{nc}%
\end{align}
where $V_{0}$ is the averaged scattering strength, $n_{im}$\ is the impurity
density, $\left(  \mathbf{k\times k}^{\prime}\right)  _{\mu\nu}\equiv k_{\mu
}k_{\nu}^{\prime}-k_{\nu}k_{\mu}^{\prime}$. Thus an interband off-shell
scattering ($\eta^{\prime\prime\prime}\neq\eta$) is unavoidable in each term
of the noncrossing intrinsic-skew-scattering \cite{Sinitsyn2007,Kovalev2010}.
More specifically, in Eq. (\ref{nc}) for $\omega_{ll^{\prime}}^{4a-nc}$, the
contribution related to $\delta_{\mathbf{k}^{\prime\prime}\mathbf{k}%
^{\prime\prime\prime}}$ ($\delta_{\mathbf{k}^{\prime}\mathbf{k}^{\prime
\prime\prime}}+\delta_{\mathbf{kk}^{\prime\prime\prime}}$) corresponds to the
sum of the middle parts of the two (four) Feynman diagrams in the last row
(first and second rows) of Fig. 11 in Ref. \cite{Sinitsyn2007}.\

For the crossed contribution, we get
\begin{align}
\omega_{ll^{\prime}}^{4a-c}  &  =-\frac{\left(  2\pi n_{im}V_{0}^{2}\right)
^{2}}{2\hbar}\delta_{\eta^{\prime}\eta}\delta\left(  d_{ll^{\prime}}\right)
\sum_{l^{\prime\prime},l^{\prime\prime\prime}}^{\prime}\frac{\delta
_{\eta^{\prime\prime}\eta}\delta\left(  d_{ll^{\prime\prime}}\right)
}{d_{ll^{\prime\prime\prime}}}\nonumber\\
&  \times\left(  \mathbf{k\times k}^{\prime}+\mathbf{k}^{\prime}%
\times\mathbf{k}^{\prime\prime}+\mathbf{k}^{\prime\prime}\times\mathbf{k}%
\right)  _{\mu\nu}\nonumber\\
&  \times\left(  \delta_{\mathbf{k+\mathbf{k}}^{\prime}\mathbf{\mathbf{=k}%
}^{\prime\prime}\mathbf{+k}^{\prime\prime\prime}}+\delta_{\mathbf{k+\mathbf{k}%
}^{\prime\prime}\mathbf{\mathbf{=k}}^{\prime}\mathbf{+k}^{\prime\prime\prime}%
}+\delta_{\mathbf{k+\mathbf{k}}^{\prime\prime\prime}\mathbf{\mathbf{=k}%
}^{\prime\prime}\mathbf{+\mathbf{k}}^{\prime}}\right) \nonumber\\
&  \times\left\{  \mathrm{Im}\left[  J_{\mu}^{\eta\eta^{\prime\prime\prime}%
}\left(  \mathbf{k}\right)  J_{\nu}^{\eta^{\prime\prime\prime}\eta}\left(
\mathbf{k}\right)  \right]  -\delta_{\eta^{\prime\prime\prime}\eta}\Omega
_{\mu\nu}\left(  \mathbf{k}\right)  \right\}  , \label{c}%
\end{align}
which contains both intraband ($\eta^{\prime\prime\prime}=\eta$) and interband
($\eta^{\prime\prime\prime}\neq\eta$) terms. $\Omega_{\mu\nu}\left(
\mathbf{k}\right)  $ is the momentum-space Berry curvature. We note that Eq.
(\ref{c}) was already obtained by Luttinger sixty years ago
\cite{Luttinger1958}, but has been unnoticed \cite{Sinitsyn2006,Sinitsyn2008}
in the recent researches on the crossed contribution to anomalous and spin
Hall effects. But Luttinger missed the noncrossing term $\omega_{ll^{\prime}%
}^{4a-nc}$. In Eq. (\ref{c}), the contributions related to $\delta
_{\mathbf{k+\mathbf{k}}^{\prime}\mathbf{\mathbf{=k}}^{\prime\prime}%
\mathbf{+k}^{\prime\prime\prime}}$ and $\delta_{\mathbf{k+\mathbf{k}}%
^{\prime\prime}\mathbf{\mathbf{=k}}^{\prime}\mathbf{+k}^{\prime\prime\prime}%
}+\delta_{\mathbf{k+\mathbf{k}}^{\prime\prime\prime}\mathbf{\mathbf{=k}%
}^{\prime\prime}\mathbf{+\mathbf{k}}^{\prime}}$ correspond to the middle parts
of the X diagram and two $\Psi$ diagrams \cite{Ado2015,Ado2016,Ado2017},
respectively. The presence of crossed intrinsic-skew-scattering reveals the
fact that, not only interband off-shell processes but also intraband off-shell
processes are important for anomalous and spin Hall effects as well as related phenomena.

\section{Discussion}

We conclude by discussing certain limitations of the present consideration.

In the case of weak disorder-potential, the off-diagonal response only
concerns the lowest nonzero order of $\left\langle f_{ll^{\prime}%
}\right\rangle $, while the analysis of $\left\langle f_{l}\right\rangle $ has
to go to higher orders in the perturbation expansion in terms of the disorder
potential. Because this expansion is basically a bare perturbation theory,
some trivial renormalization terms are unavoidable in high orders of this
expansion. These terms should be eliminated systematically by a
renormalization procedure, if one aims at placing the density matrix
equation-of-motion theory as a generic foundation for transport in the entire
metallic region. Such a renormalization treatment has been shown for
non-relativistic free electrons \cite{Moore1967,Watabe1966}, but has never
been done for Bloch electrons according to our literature knowledge. On the
other hand, we only regard the density matrix approach as a foundation of the
Boltzmann theory in the case of very weak disorder potential. In this case the
aforementioned trivial renormalization effects are much smaller high-order
corrections (at least $O\left(  \lambda^{2}\right)  $ for spin-orbit induced
transport coefficients), and can thus be neglected.

The present consideration shows that the modified Boltzmann formalism
\cite{Sinitsyn2006,Sinitsyn2007,Sinitsyn2008,Xiao2017SOT-SBE} only works in
the presence of weak disorder-potentials. This is much more limited than the
usually accepted regime of (qualitative) validity of a Boltzmann theory
$\hbar/\tau<\Delta$ (this is a necessary condition for the validity of the
Kohn-Luttinger expansion), where $\tau$ is the electron lifetime and $\Delta$
is the minimal interband-splitting around the Fermi level
\cite{Tanaka2008,Xiao2017SOT}. This implies that there exists another sort of
modified Boltzmann formalism in the presence of very dilute impurities,
because large $\tau$ can be obtained by not only weak disorder potential but
also small impurity density. In fact this kind of theory also started from
Luttinger and Kohn \cite{LK1958} who tried to formulate the transport equation
according to the order of impurity density. But this Boltzmann theory is
complicated enough because it is nonperturbative with respect to the disorder
potential, especially when concerning the distribution function in the zeroth
order of impurity density \cite{LK1958}. In particular, this nonperturbative
nature indicates that the Gaussian disorder approximation cannot well describe
dilute impurities with strong scattering potentials. New characteristics of
transport, differing from those discussed above in moderately dirty samples
with weak disorder-potentials, are naturally expected in the regime of dilute
impurity and strong disorder-potential (as revealed in recent Kubo
diagrammatic calculations \cite{Milletari2016RC}).

\begin{acknowledgments}
We acknowledge very insightful discussions with Chunli Huang and Qian Niu. C.X. and B.X. are supported by DOE (DE-FG03-02ER45958, Division of Materials Science and Engineering), NSF (EFMA-1641101) and Welch Foundation (F-1255).
J.Z. is supported by the Welch Foundation under Grant No. TBF1473. The analysis in Sec. II is supported by the DOE grant.
\end{acknowledgments}

\appendix

\section{Calculation details}

\begin{widetext}%
The expressions for $\omega_{l^{\prime}l}^{\left(  4\right)  }$ and
$\omega_{ll^{\prime}}^{\left(  4\right)  }$ are
\begin{align}
\omega_{l^{\prime}l}^{\left(  4\right)  } &  =\frac{2\pi}{\hbar}%
\sum_{l^{\prime\prime},l^{\prime\prime\prime}}^{\prime}\delta\left(
d_{ll^{\prime}}\right)  \left[  \frac{\left\langle H_{ll^{\prime\prime\prime}%
}^{^{\prime}}H_{l^{\prime\prime\prime}l^{\prime}}^{^{\prime}}H_{l^{\prime
}l^{\prime\prime}}^{^{\prime}}H_{l^{\prime\prime}l}^{^{\prime}}\right\rangle
}{d_{ll^{\prime\prime}}^{+}d_{ll^{\prime\prime\prime}}^{-}}+\frac{\left\langle
H_{ll^{\prime\prime\prime}}^{^{\prime}}H_{l^{\prime\prime\prime}%
l^{\prime\prime}}^{^{\prime}}H_{l^{\prime\prime}l^{\prime}}^{^{\prime}%
}H_{l^{\prime}l}^{^{\prime}}\right\rangle }{d_{ll^{\prime\prime}}%
^{-}d_{ll^{\prime\prime\prime}}^{-}}+\frac{\left\langle H_{ll^{\prime}%
}^{^{\prime}}H_{l^{\prime}l^{\prime\prime}}^{^{\prime}}H_{l^{\prime\prime
}l^{\prime\prime\prime}}^{^{\prime}}H_{l^{\prime\prime\prime}l}^{^{\prime}%
}\right\rangle }{d_{ll^{\prime\prime}}^{+}d_{ll^{\prime\prime\prime}}^{+}%
}\right]  ,\nonumber\\
\omega_{ll^{\prime}}^{\left(  4\right)  } &  =\frac{2\pi}{\hbar}%
\sum_{l^{\prime\prime},l^{\prime\prime\prime}}^{^{\prime}}\delta\left(
d_{ll^{\prime}}\right)  \left[  \frac{\left\langle H_{ll^{\prime\prime\prime}%
}^{^{\prime}}H_{l^{\prime\prime\prime}l^{\prime}}^{^{\prime}}H_{l^{\prime
}l^{\prime\prime}}^{^{\prime}}H_{l^{\prime\prime}l}^{^{\prime}}\right\rangle
}{d_{ll^{\prime\prime}}^{-}d_{ll^{\prime\prime\prime}}^{+}}+\frac{\left\langle
H_{ll^{\prime}}^{^{\prime}}H_{l^{\prime}l^{\prime\prime}}^{^{\prime}%
}H_{l^{\prime\prime}l^{\prime\prime\prime}}^{^{\prime}}H_{l^{\prime
\prime\prime}l}^{^{\prime}}\right\rangle }{d_{ll^{\prime\prime}}%
^{-}d_{ll^{\prime\prime\prime}}^{-}}+\frac{\left\langle H_{ll^{\prime
\prime\prime}}^{^{\prime}}H_{l^{\prime\prime\prime}l^{\prime\prime}}%
^{^{\prime}}H_{l^{\prime\prime}l^{\prime}}^{^{\prime}}H_{l^{\prime}%
l}^{^{\prime}}\right\rangle }{d_{ll^{\prime\prime}}^{+}d_{ll^{\prime
\prime\prime}}^{+}}\right]  .\label{scattering-rate-4}%
\end{align}
In fact, obtaining this equation is not easy, if one does not expect Eq.
(\ref{golden-rule}) in advance.
Starting from the expressions for $C_{l}^{^{\prime\prime}}$, $C_{ll^{^{\prime
}}}^{\left(  0\right)  }$, $C_{ll^{^{\prime}}}^{\left(  1\right)  }$ and
$C_{l}^{\left(  2\right)  }$ presented in the main text,\ after some delicate
steps we recover Luttinger's expression \cite{Luttinger1958} for
$C_{l}^{^{\prime\prime}}$:%
\[
C_{l}^{^{\prime\prime}}=ie\mathbf{E\cdot}\left\{  \frac{\partial}%
{\partial\mathbf{k}}\rho_{ll}^{\left(  2\right)  }+\sum_{l^{\prime}}%
^{^{\prime}}\left[  \frac{\left\langle H_{l^{\prime}l}^{^{\prime}}%
\mathbf{\hat{D}}H_{ll^{\prime}}^{^{\prime}}\right\rangle +\left\langle
\left\vert H_{ll^{\prime}}^{^{\prime}}\right\vert ^{2}\right\rangle \left(
\mathbf{J}_{l}-\mathbf{J}_{l^{\prime}}\right)  }{d_{ll^{\prime}}^{-}}%
\frac{\rho_{l}-\rho_{l^{\prime}}}{\epsilon_{l}-\epsilon_{l^{\prime}}}%
+\frac{\left\langle \left\vert H_{ll^{\prime}}^{^{\prime}}\right\vert
^{2}\right\rangle }{d_{ll^{\prime}}^{-}}\mathbf{\hat{D}}\frac{\rho_{l}%
-\rho_{l^{\prime}}}{\epsilon_{l}-\epsilon_{l^{\prime}}}+c.c.\right]  \right\}
,
\]
which can also be expressed as
\[
C_{l}^{^{\prime\prime}}=i\hbar e\mathbf{E\cdot}\sum_{l^{\prime}}^{^{\prime}%
}\omega_{ll^{\prime}}^{\left(  2\right)  }\delta\mathbf{r}_{l^{\prime}%
l}\left(  -\frac{\partial\rho_{l}}{\partial\epsilon_{l}}\right)
+ie\mathbf{E\cdot}\left\{  \frac{\partial\rho_{ll}^{\left(  2\right)  }%
}{\partial\mathbf{k}}+\sum_{l^{\prime}}^{^{\prime}}\left[  \frac
{\mathbf{\hat{D}}\left\langle \left\vert H_{ll^{\prime}}^{^{\prime}%
}\right\vert ^{2}\right\rangle }{\epsilon_{l}-\epsilon_{l^{\prime}}}\frac
{\rho_{l}-\rho_{l^{\prime}}}{\epsilon_{l}-\epsilon_{l^{\prime}}}%
+\frac{2\left\langle \left\vert H_{ll^{\prime}}^{^{\prime}}\right\vert
^{2}\right\rangle }{\epsilon_{l}-\epsilon_{l^{\prime}}}\mathbf{\hat{D}}%
\frac{\rho_{l}-\rho_{l^{\prime}}}{\epsilon_{l}-\epsilon_{l^{\prime}}}\right]
\right\}  .
\]
By using $\rho_{ll}^{\left(  2\right)  }=\sum_{l^{\prime}}^{^{\prime}}%
\frac{\left\vert H_{ll^{\prime}}^{^{\prime}}\right\vert ^{2}}{\epsilon
_{l}-\epsilon_{l^{\prime}}}\left[  \frac{\partial\rho_{l}}{\partial
\epsilon_{l}}-\frac{\rho_{l}-\rho_{l^{\prime}}}{\epsilon_{l}-\epsilon
_{l^{\prime}}}\right]  =\epsilon_{l}^{\left(  2\right)  }\frac{\partial
\rho_{l}}{\partial\epsilon_{l}}-\sum_{l^{\prime}}^{^{\prime}}\frac{\left\vert
H_{ll^{\prime}}^{^{\prime}}\right\vert ^{2}}{\epsilon_{l}-\epsilon_{l^{\prime
}}}\frac{\rho_{l}-\rho_{l^{\prime}}}{\epsilon_{l}-\epsilon_{l^{\prime}}}$,
$C_{l}^{^{\prime\prime}}$ can be cast into%
\[
C_{l}^{^{\prime\prime}}=i\hbar e\mathbf{E\cdot}\sum_{l^{\prime}}^{^{\prime}%
}\omega_{ll^{\prime}}^{\left(  2\right)  }\delta\mathbf{r}_{l^{\prime}%
l}\left(  -\frac{\partial\rho_{l}}{\partial\epsilon_{l}}\right)  +i\hbar
e\mathbf{E\cdot}\left\{  \left[  \frac{\partial\epsilon_{l}^{\left(  2\right)
}}{\hbar\partial\mathbf{k}}+\sum_{l^{\prime}}^{^{\prime}}\frac{\left\langle
\left\vert H_{ll^{\prime}}^{^{\prime}}\right\vert ^{2}\right\rangle }{\left(
\epsilon_{l}-\epsilon_{l^{\prime}}\right)  ^{2}}\mathbf{v}_{l}^{0}\right]
\frac{\partial\rho_{l}}{\partial\epsilon_{l}}+\mathbf{v}_{l}^{0}\epsilon
_{l}^{\left(  2\right)  }\frac{\partial^{2}\rho_{l}}{\partial\epsilon_{l}^{2}%
}-\sum_{l^{\prime}}^{^{\prime}}\frac{\mathbf{v}_{l^{\prime}}^{0}\left\langle
\left\vert H_{ll^{\prime}}^{^{\prime}}\right\vert ^{2}\right\rangle }{\left(
\epsilon_{l}-\epsilon_{l^{\prime}}\right)  ^{2}}\frac{\partial\rho_{l^{\prime
}}}{\partial\epsilon_{l^{\prime}}}\right\}  .
\]
The second term on the right hand side still exists in the case of
non-relativistic free electrons without spin-orbit coupling. It contains some
trivial renormalization effects (e.g., $\epsilon_{l}^{\left(  2\right)  }$ is
the second-order energy correction in the bare quantum mechanical perturbation
theory) and is not important to spin-orbit induced transport. Thus we only
preserve $C_{l}^{^{\prime\prime}}=i\hbar e\mathbf{E\cdot}\sum_{l^{\prime}%
}^{^{\prime}}\omega_{ll^{\prime}}^{\left(  2\right)  }\delta\mathbf{r}%
_{l^{\prime}l}\left(  -\frac{\partial\rho_{l}}{\partial\epsilon_{l}}\right)  $
which is of qualitative importance for spin-orbit induced transport
\cite{Xiao2017SOT-SBE,Sinitsyn2007,Sinitsyn2008}.
\end{widetext}


\begin{thebibliography}{99}
\bibitem {note-SOscattering}In the present paper we do not consider spin-orbit
interaction around the impurities, which can be included into the present
framework (Ref. \cite{Xiao2018JPCM}).
\end{thebibliography}
\end{document}